\documentclass[final,1p,times]{elsarticle}

\usepackage{amsthm}
\usepackage{amsmath}

\journal{arxiv.org}

\newcommand{\Zed}{\mathbb Z}

\newcommand{\Real}{\mathbb R}

\newtheorem{theorem}{Theorem}

\newtheorem{lemma}{Lemma}
\newtheorem{corollary}[theorem]{Corollary}
\newdefinition{remark}{Remark}
\newdefinition{example}{Example}
\newdefinition{definition}{Definition}

\begin{document}

\begin{frontmatter}



\title{A state of a dynamic computational structure distributed in an environment: a model and its corollaries}


\author{Oleksiy Kurganskyy}
\ead{kurgansk@gmx.de}
\address{Institute of Applied Mathematics and Mechanics, Ukrainian National Academy of Sciences, 74 R. Luxemburg St, Donetsk, Ukraine}

\begin{abstract}
In this work a collective of interacting stateless automata in a discrete geometric environment is considered as an integral automata-like computational dynamic object. For such distributed on the environment object different approaches to definition of the measure of state transition are possible. We propose an approach for defining what a state is.
\end{abstract}


\end{frontmatter}


\section{Introduction}

Currently there is great interest in computational models consisting of underlying regular computational environments, and built on them distributed computational structures. Examples of such models are cellular automata, spatial computation and space-time crystallography~\cite{pedestrian}. For any computational model it is natural to define a functional equivalence of different but related computational structures. In the finite automata theory an example of such equivalence is automata homomorphism and, in particular, automata isomorphism. If we continue to stick to the finite automata theory, a fundamental question arise, what a state of a distributed computational structure is. This work is devoted to particular solution of the issue.

The work consists of the informal presentation in this introduction of an idea that came from the Poincar\'e's relativity theory, and an illustration of this idea by a simple computational model with a regular and discrete dynamics that is especially suited for the illustration purpose. The model with the problem statement is like the model of~\cite{pedestrian}, but in essence differs from it. One of the distinguishing features of the model is spatial movement of computational structure in the environment. As consequence we present the number of results on the relationship between computational and dynamic properties of these structures.

So, in this work the collectives of stateless (i.e. with one state) automata interacting with an environment defined as a graph are considered. We study a collective of automata as an integral automata-like dynamic computational object. The fundamental question what is the state of such dispersed and moving on the environment object and how to measure the amount of state transitions is quite non-trivial. As opposed to the finite state automata where the measure of state transition is one state per unit of time, for a computational dynamic object distributed on the environment certainly different approaches to definition of the measure of state transition are possible. 

The idea of our approach came from the special relativity theory. It is based on the concept of relativity in Poincar\'e's interpretation ~\cite{poincare}. In explanation how to generally understand the relativity Poincar\'e begins with an example of resizing of all dimensions in the Universe in the same number of times and proceeds with consideration of arbitrary deformations concluding that they should be unnoticed by an observer because observer’s standards are subjected to the same deformations. This reasoning coupled with the principle that without any changes in the object the process of computation in it is not possible is used in this study to define a state of collective automata. A ``change'' in an object is a change in the relative position of its ``elementary'' parts per se. Thus, the movement in the environment underlies the process of computation in our model.

Let us explain it by an example of a chessboard with pawns. The chessboard is provided with a natural reference frame. Suppose that we can move any pawn one chess square per unit time in one of four directions: $\leftarrow$,$\uparrow$,$\rightarrow$,$\downarrow$, i.e. pawn's velocity is one chess square in a certain directions per unit time. Let us compose from the pawns a figure, for example, an ``O''-like figure, and look at all these pawns as an integral object. Define the velocity of the object on the chessboard as the average velocity of his pawns. Suppose that the object is moved at maximal velocity ``one chess square per unit time'' in a constant direction. Can the object be transformed simultaneously with the motion from ``O'' to, for example, ``T''? It is obviously that no.  That is, at maximum constant velocity in the example the object cannot be changed and, from our point of view, its state is invariable and it performs no computation. This point of view we have formally illustrated in this work by the simplest example model of stateless automata interacting with one-dimensional environment. 

The introduced illustrative model is computational universal, and collectives of automata in the environment can be seen as automata-like computational objects. By analogy with Turing machines, which can answer certain questions about properties of words on the tapes, for these objects natural questions arise what properties of the environment and other objects in it they can identify. One of the interesting questions is what can an object say about the velocity of its elementary parts (i.e. stateless automata). Can it ``perceive'' any changes in velocity of elementary parts which it consists of? This question is similar to the issue in the Poincar\'e's story about relativity: can the observer see the deformation of the space, which includes the deformation measurement standards? Having as a goal the answer ``no'', we define our computational model. This goal determines the language (motion velocity, proper time velocity as a measure of state transition, reference frame) of interaction between collectives of automata. 

The concluding comparison of the obtained results with some formulas of special relativity theory shows that the formulated principles are invariant in relation to physical and informational linguistic means of expression. In other words, the semantic affinity of the original principle of motion in our discrete model to the principles of the special relativity theory resulted in the syntactic affinity of their languages (e.g., time dilation formula, velocity-addition formula, ``length contraction/extension'' formula).  But because of discreteness of our model there are differences. For example, the linear sizes of objects can either decrease or increase in different reference frames. Comparing the formulas with the formulas of the special relativity theory allows also revealing different physical meaning of Lorentz factor in the formulas of length contraction and time dilation. 

To emphasize a physical analogy in the proposed model and the problem statement we use the short word ``body'' as alias for ``collective of automata''.

The proposed research is inspired by several research directions which are: 1) collective of automata in finite automata theory and complexity of the interactions between them; 2) discrete models of physical processes and projecting of the physical world into informational space of symbols and languages for computer simulation of the physical world; 3) studying the notion of time; 4) spatial computation.

\begin{figure}\label{shess}
\begin{center}
\includegraphics[scale=0.3]{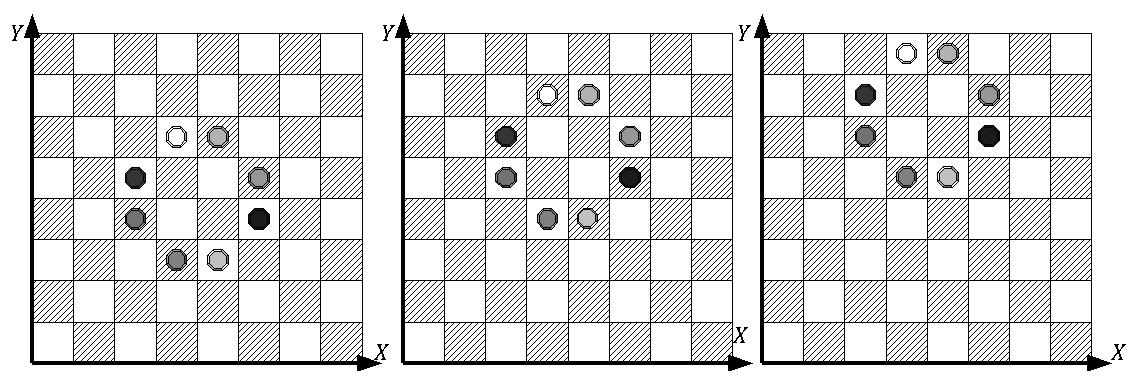}
\caption{A chessboard with pawns}
\end{center}
\end{figure}

\section{Definitions}\label{body}

In what follows we use denotations $\Zed$ and $\Real$ for the sets of integers and real numbers, respectively. 
Initially in the model defintion we assume that the domains for the time $T$ and space coordinates $X$ coincide with $\Zed$ 
but then we will extend them to $\Real$. 

The computational model, that we use for this study, 
consists of two main components: an underlying environment $G$ that is represented by a graph and a set of stateless automata, which are interacting with the environment.

\begin{figure}\label{environ}
\begin{center}
\includegraphics[scale=0.4]{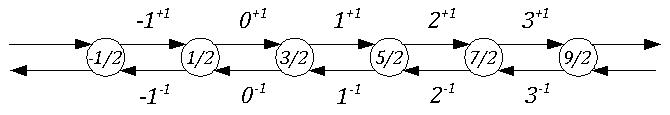}
\caption{The environment $G$}
\end{center}
\end{figure}

The environment $G$ (see Fig.\ref{environ}) is defined as the infinite directed graph with the set of nodes $V=\{x+\frac{1}{2} | x \in \Zed \}$ and the set of edges $E= \{ (x-\frac{i}{2},x+\frac{i}{2}) | x \in \Zed, i \in \{-1,1\} \}$. An edge $(x-\frac{i}{2},x+\frac{i}{2})$ for some $i \in \{-1,1\}$ has the absolute coordinate $x\in \Zed$ and the direction $i$. Absolute coordinate of an edge $e$ will be denoted by $x(e)$ and its direction by  $r(e)$. Also the edge $e$ will be denoted by $x(e)^{r(e)}$. 
By the neighborhood of an edge $x^i$ we understand the pair of edges $x^i$ and $(x+i)^{-i}$. The edges $x^i$ and $(x+i)^{-i}$ will be called opposite edges and $x^i$ and $x^{-i}$ will be called contrary edges.

Let $A=(S_A,I_A,O_A,{\delta}_A, {\lambda}_A)$ be a Mealy automaton, where $S_A$, $I_A$ and $O_A$ are the sets of states,
input symbols, and output symbols, respectively, and
${\delta}_A: S_A \times I_A \rightarrow S_A$
and ${\lambda}_A: S_A \times I_A \rightarrow O_A$  are transition function and
function of outputs
respectively. We consider only stateless Mealy automata. The set of states of a stateless automaton $A$ consists of single state, so there is no sense to mention the transition function and the set of states of a stateless automaton. Thus we write $A=(I_A,O_A,\lambda_A)$ instead of $A=(S_A,I_A,O_A,{\delta}_A, {\lambda}_A)$.  Within the framework of this article for reasons of consistency of latter definitions we name a stateless Mealy automaton as an {\bf elementary body}, and we name its unique state as the {\bf internal state}. The elementary bodies will be denoted by lowercase letters, for example, $b=(I_b,O_b,\lambda_b)$. We assume also that elementary bodies are coloured in a way that isomorphic automata will have the same colour and non-isomorphic automata will have different colours. We assume that $r$ different numbered from $1$ to $r$ colours are used. Every moment of time $t$ any elementary body $b$ is located on an edge $b(t)$ of the environment $G$. 
The input for an elementary body, located on an environment edge $x^i$, is the sequence $(p_1,p_2,\ldots,p_r,q_1,q_2,\ldots,q_r)\in I_b$ called the {\bf neighbourhood state} of the edge $x^i$, where $p_k$ and $q_k$ are the numbers of elementary bodies of the colour $k$, 
located on the edges $x^i$ and $(x+i)^{-i}$ at the same moment of time, respectively.
The output of an elementary body is one of the two motions either the straight-line motion $\rightarrow\in O_b$ or the turn $\hookleftarrow\in O_b$. 
If the output of an elementary body $b$ at a time moment $t\in\Zed$ on an edge $b(t)=x^i$ is the straight-line motion, then at the next time moment $b(t+1)=(x+i)^i$ and we say that it does not change its {\bf external state}. If the output is the turn then $b(t+1)=x^{-i}$ and we say that the elementary body changes its {\bf external state}. 

Accordint to definition all elementary bodies have the same sets of input and output symbols, so we can write $b=(I,O,\lambda_b)$ instead of $b=(I_b,O_b,\lambda_b)$.

\begin{figure}\label{euclid}
\begin{center}
\includegraphics[scale=0.4]{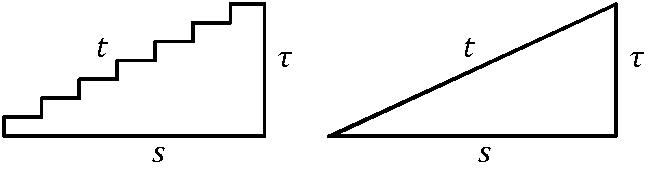}
\caption{$t=s+\tau$ vs. $t^2=s^2+\tau^2$}
\end{center}
\end{figure}

Denoting by $\tau_b(t)$ the number of external state changes of $b$ until the moment of time $t$ we have that $1=\tau_b(t+1)-\tau_b(t)+\left|x(b(t+1))-x(b(t))\right|$ and also $t=\tau_b(t)-\tau_b(0)+s_b(t)$, where $s_b(t)= \sum^{t}_{j=1}|x(b(j))-x(b(j-1))|$ is the path covered by $b$ during the period of time from $0$ to $t$. 
In other words any elementary body uses the absolute time unit $t$ 
either for one spatial coordinate change $s$ in the environment or for one external state transition $\tau$. 
Schematically, this has the form $t=s + \tau$, illustrated in the left side of Figure~\ref{euclid}. On the right in Figure~\ref{euclid}, for comparison, a similar formula $t^2=s^2+\tau^2$ is shown that one would expect in a space with Euclidean metric.
This formula is using in the special relativity theory for the expression of so-called spacetime interval $\tau^2=s^2-t^2$, which is invariant under Lorentz transformations.

We call $\tau = \tau_b(t)$ the {\bf proper time} of $b$ and $w_b(t)=\tau_b(t+1)-\tau_b(t)$ the {\bf proper time velocity} of $b$. We call $w_b(t)$ uniform proper time velocity if $w_b(t)$ is a constant. 
Let us denote by $x_b(t)=x(b(t))$. We call $x_b(t)$ the {\bf absolute coordinate} of $b$ at the moment of time $t$.
We denote by $v_b(t)=x_b(t+1)-x_b(t)$ the {\bf absolute spatial velocity} of $b$ at the moment of time $t$. We call it uniform spatial velocity if $v_b(t)$ is a constant. For example, it follows from above definitions that any elementary body can have only one of the following uniform spatial velocities: $v=1$, $v=-1$, $v=0$.

An elementary body is unambiguously definable by the set of input symbols that change its external state. So in the context of our model we can understand under an elementary body $b$ the set of signals from $I$ that change its external state rather than the triple $b=(I,O,\lambda_b)$. In additional we assume also that elementary body can not change its external state anyway if opposite environment edge is empty, that is, the interaction between elementary bodies occurs only by collisions in the vertices of the environment (compare with the notion of vacuum state in~\cite{pedestrian}). The elementary bodies can be seen as analogues of signals propagating in the causal network~\cite{pedestrian}. Propagation of signals in~\cite{pedestrian} depends on the functions in the nodes of a causal network, in our model it depends on the output functions $\lambda$ of elementary bodies, i.e., on the properties of ``signal''.

It should be noted that a set $I$ can be infinite. We have done nothing to circumvent this problem but we can simply assume that the interaction of elementary bodies proceeds so that the set of all possible input symbols can only be finite.


We call the pair of a space coordinate $x$ and a time coordinate $t$ as coordinate in the 
absolute reference frame $O=X \times T $ and denote by the column vector. We call $O$ also the event space.

\begin{definition}
A body is an arbitrary finite set of elementary bodies.
\end{definition}

According to the defintion different bodies may have common parts and one body can contain another body as a subset.
If an elementary body belongs to a body then we will look at it as an elementary part of this body. An elementary body can be an elementary part of different bodies simultaneously.

The following two examples illustrate some of introduced definitions. 
Any elementary body in both examples changes its external state if and only if opposite environment edge is not empty.
From it follows that all elemantary bodies are automata isomorphic.
We assume that all elementary bodies in each example are enumerated by integer numbers. 

\begin{example}\label{example1}
At time $t=0$ for each $x \in \Zed$ the elementary body with the number $x$ is located on the edge $x^{+1}$ if $x$ is even number and $x^{-1}$ otherwise. We define the body $A_1$ as the set $A_1=\{0,1,2\}$ of elementary bodies $0$, $1$ and $2$. 
\begin{figure}
\begin{center}
\includegraphics[scale=0.4]{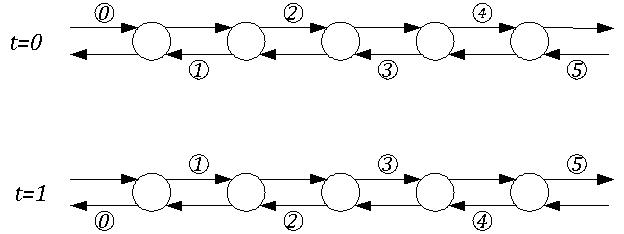}
\caption{Dynamics of elementary bodies from the Example~\ref{example1}}
\end{center}
\end{figure}
\end{example}

\begin{example}\label{example2}
At time $t=0$ for each $x \in \Zed$
the elementary body with the number $x$ has the coordinate $4 \left\lfloor \frac{x}{3} \right\rfloor+ (x \mod 3)$ and located on the edge with the direction $-1$ if $x \equiv 1 \mod 3$ and on the edge with the direction $+1$ otherwise. In this example we define the body $A_2=\{0,1,2\}$.
\end{example}

Let a body $B$ consist of $n$ elementary bodies enumerated by numbers 
$\{1,2, \ldots, n\}$. Then the absolute (average) coordinate of the body $B$ at time $t$ is the value
$x_B(t)=\frac{x_1(t)+ \ldots + x_n(t)}{n}$ and absolute spatial velocity of the body $B$ at time $t$ is the value 
$v_B(t)=x_B(t+1)-x_B(t)$. The bodies $A_1$ and $A_2$ from the above examples have 
uniform spatial velocities $0$ and $\frac{1}{3}$, respectively. From the definitions it follows that the
maximal possible positive or negative spatial velocities of any body can be $1$ or $-1$.

Since the coordinate values of a body can be non-integers let us
extend the absolute reference frame $O$ from $\Zed \times \Zed$ to $\Real \times \Real$.  
Let $t \in \Zed$ and $-\frac{1}{2}< \Delta \leq \frac{1}{2}$ then we say that an elementary body $b$  
at time $t+\Delta$ has the coordinate $x_b(t+\Delta)=x_b(t)+r(b(t))\cdot\Delta$ and is located on the edge 
$b(t+ \Delta)=b(t)$. 
The definition implies that the all neighborhood states of the environment edge $b(t')$ at time $t'=t+\Delta$ coincide for all $\Delta$, $-1/2<\Delta\le 1/2$, $t\in\Zed$.
In particular, the neighborhood state of the environment edge $b(t')$ at times $t'=t$ and $t'=t+1/2$ are the same, $t\in\Zed$, and thus, the behavior of elementary bodies is completely determined in the nodes of the environment in which, figuratively speaking, the elementary bodies collide.

We define the world line $b(t':t'')$ of an elementary body $b$ in time interval from $t'\in\Real$ to $t''\in\Real$ as $b(t':t'')=\left\{\left( \begin{array}{c}
x_b(t)\\t\end{array} \right)\left| t' \le t \le t'', t \in \Real\right. \right\}$. 

If the motion of an elementary body in a time interval from $t'$ to $t''$ corresponds to a straight world line segment, that is, during this time the elementary body did not change the external state, such a motion is called {\bf elementary motion}.

\section{A state of a body}\label{state}

A body interacting with other bodies exert influence on them and is also under their
influence. It is quite natural to describe such influences on the basis of the notion of a state of a body. Our definition of a state of a body takes into consideration the relative positioning of its elementary parts in the environment. The changes of relative positioning of elementary parts in a body can affect the body entirely or a particular part of it. This motivates the question how to measure the amount of state transition. Before the definition of the notion of a state we introduce the denotation for the measure $\tau = \tau_B(t)$ of state transition of a body $B$ with the flow of time $t$. A casual meaning of $\tau = \tau_B(t)$ is the 
``age'' of the body $B$ at the moment $t$. We call $\tau = \tau_B(t)$ the proper time of $B$.

Independently from the definition of $\tau = \tau_B(t)$, we introduce the velocity $w_B(t)$ of {\bf external state} transition of the body $B$ as $w_B(t)=\tau_B(t+1)-\tau_B(t)$. We call this value as the proper time velocity of $B$ at the moment of the absolute time $t$.

\begin{definition}
For any body $B$ $w_B(t)=1 \Leftrightarrow \forall_{b \in B}w_b(t)=1$
\end{definition}

\begin{definition}
For any body $B$ $w_B(t)=0 \Leftrightarrow \forall_{b \in B}w_b(t)=0$
\end{definition}
From it follows that a body $B$ does not change its {\bf external state} 
if all its elementary bodies do not change their external states. It means that two bodies are at the same external state in the environment if one of them can be transformed into another by 
straight-line shifts on the equal number of steps applied to all its elementary parts in direction corresponding their external states.

We have defined what does it mean that two bodies are in the same external state, rather than what the external state of a body in fact is. If needed the notion of external state can be in generally defined as follows. Since the relation ``to be in the same external state'' is an equivalence relation, the external states are equivalence classes of this relation. The same holds for the latter definition of internal state.

\begin{theorem}
For any body $B$, if $|v_B(t)|=1$ then $w_B(t)=0$.
\end{theorem}
\begin{proof}
The statement follows from the fact that any change of the external state of a body 
is not possible in case of maximal spatial velocity of all its elementary parts.
\end{proof}

The notion of external state of a body allows to start to consider the bodies as an automata-like model of algorithms. 
It is natural to ask a functional equivalence of different bodies for example something like automata isomorphism in the finite automata theory.

But since two bodies with different absolute spatial velocities are definitely in different external states we can not compare them functionally. For example there is no sense to ``ask'' a body to determine its absolute spatial velocity. 
However we would like to identify two bodies as the same algorithm even if they move with different spatial velocities.
It will be achieved by introduction of affine isomorphism of bodies through definition of inertial reference frame associated with a body so that the external state of a body will be presented as pair of components: spatial velocity of the body and its spatial velocity invariant internal state. 
The point of introducing the notion of inertial  reference frame 
associated with a body lies in the ability to consider other bodies in 
relation to the given one. With reference frames we attempt to develop a language of interaction between bodies just as the input and output alphabets of finite Mealy automata are for the interaction between them.
The language that we develop is one of the possible and thus our approach reflects a Poincar\'e's conventional point of view on the physical laws.

An example of inertial reference frame is the absolute reference frame $O$ associated with an immovable body $B$ such that for all $t\in\Zed$ $x_B(t)=0$, $v_B(t)=0$, $w_B(t)=1$, $\tau_B(0)=0$, and, hence, $\tau_B(t)=t$. Thus, the introduced notions of absolute time, absolute coordinate and absolute spatial velocity implicitly mean an absolutely motionless body in relation to which objects were considered. The reference frames associated with the bodies allow us to make these notions relative.

Let us denote (for a pair of bodies $A$ and $B$) by $x_{AB}(\tau_B)$, $v_{AB}(\tau_B)$, $w_{AB}(\tau_B)$ and $\tau_{AB}(\tau_B)$ the coordinate, the spatial velocity, the proper time velocity and the proper time of the body $A$ at the moment of time $\tau_{B}$ in the reference frame $O_B$ associated with the body $B$, respectively.
By definition we assume that $x_{BB}(\tau_B)\equiv 0$, $v_{BB}(\tau_B)\equiv 0$, $w_{BB}(\tau_B)\equiv 1$ and $\tau_{BB}(\tau_B)=\tau_B$. 

\begin{definition}
A body $B$ is called an inertial body if $v_B(t)$ and $w_B(t)$ are both constants.
\end{definition}

The property to be inertial implies uniform changes of not only spatial coordinates but also time coordinates. For the sake of simplicity consider the case only the inertial bodies.

The only restriction imposed on the inertial reference frames is the property that space-time coordinates of same events in different inertial reference frames are connected by affine transformation.
It follows that a body that is inertial in the absolute inertial reference frame is inertial in any other inertial reference frame.

\begin{remark}\label{rem1}
It follows that $\tau_{AB}(\tau_B)=\tau_{AB}(0)+\tau_B\cdot w_{AB}$ and $x_{AB}(\tau_B)=x_{AB}(0)+\tau_B\cdot v_{AB}$ for inertial bodies $A$ and $B$.
\end{remark}

For any bodies $A$ and $B$ let us denote by $L_{BA}:O_B\rightarrow O_A$ the affine mapping that connects $O_B$ and $O_A$ such that an event $(x,\tau_B)$ in $O_B$ coincides with the event $L_{BA}(x,\tau_B)$ in $O_A$.

These assumptions are sufficient to find out $L_{BA}$. Without loss of generality we assume that the origins of both reference frames $O_A$ and $O_B$ are the same: $x_{BA}(0)=0$ and $\tau_{BA}(0)=0$. Then the mapping $L_{BA}$ is linear. Let us work out the form of transformation matrix 
$L_{BA}=\left( \begin{array}{cc}
a_{11} & a_{12}  \\
a_{21} & a_{22} \end{array} \right)$. 

The direction of a vector $\bar{a}$ is the set of vectors $\{\lambda\cdot\bar{a} | \lambda>0\}$.

\begin{lemma}\label{lemma_LBA}
The vectors $\left(\begin{array}{c} 1\\1\end{array}\right)$ and $\left(\begin{array}{c} -1\\1\end{array}\right)$ are eigenvectors of the mapping $L_{BA}$.
\end{lemma}
\begin{proof}
The directions of reference frame axes are imaginary directions in the event space. But the directions of the vectors $\left(\begin{array}{c} 1\\1\end{array}\right)$ and $\left(\begin{array}{c} -1\\1\end{array}\right)$ in the absolute reference frame correspond to the two only possible directions of elementary motions of elementary bodies going from the reference frame origin and therefore they do not depend on reference frames. It follows that these directions {\sl are invariant} by any affine transformation.
\end{proof}

\begin{corollary}
For the matrix $L_{BA}$ holds $a_{11}=a_{22}$ and $a_{12}=a_{21}$.
\end{corollary}
\begin{proof}
Based on the lemma~\ref{lemma_LBA} the corollary statement follows as a result of straightforward calculations.
\end{proof}

Note that the set of directions of the vectors $\left(\begin{array}{c} 1\\1\end{array}\right)$ and $\left(\begin{array}{c} -1\\1\end{array}\right)$ is also invariant under the transformation $L_{BA}$. It will remain invariant, if we let the transformation $L_{BA}$ permute the directions. In this case we have $a_{11}=-a_{22}$ and $a_{12}=-a_{21}$. In physics for these two situations the different words are used, namely the standard and symmetric configuration. We consider only the standard configuration, that is 
$a_{11}=a_{22}$ and $a_{12} = a_{21}$, because only the standard configuration satisfies by lemma~\ref{lemma_LBA} our requirements to the inertial reference systems.

\begin{theorem}\label{theorem1} It holds
$$L_{BA}=
\left( \begin{array}{cc}
1/w_{BA} & v_{BA}/w_{BA}  \\
v_{BA}/w_{BA} & 1/w_{BA} \end{array} \right).
$$
\end{theorem}
\begin{proof}

Since 
$L_{BA} \cdot 
\left( \begin{array}{c}
x_{BB}(\tau_{BA}(\tau_A))  \\
\tau_{BA}(\tau_A)
\end{array} \right)
=
\left(\begin{array}{c}
x_{BA}(\tau_A)\\
\tau_A\end{array}\right)
$, 
$x_{BA}(\tau_A)=v_{BA}(\tau_A)\cdot\tau_A$, 
$\tau_{BA}(\tau_A)=w_{BA}(\tau_A)\cdot\tau_A$,
$x_{BB}(\tau_B) \equiv 0$, 
 then 
$L_{BA} \cdot 
\left( \begin{array}{c}
0  \\
1
\end{array} \right)
=
\left( \begin{array}{c}
v_{BA}/w_{BA}  \\
1/w_{BA} \end{array} \right)
$.
From it follows that $a_{12}=v_{BA}/w_{BA}$ and $a_{22}=1/w_{BA}$.
\end{proof}

The following corollaries hold for any intertial bodies $A$, $B$, $C$.

\begin{corollary}\label{cor3}
It holds $v_{AB}=-v_{BA}$ and $w_{AB}\cdot w_{BA}=1-v_{AB}^2=1-v_{BA}^2.$
\end{corollary}
\begin{proof}
The equalities can be derived from $L_{AB}\cdot L_{BA}= 
\left( \begin{array}{cc}
1 & 0  \\
0 & 1 \end{array} \right)
$.
\end{proof}

\begin{corollary}\label{cor_addition}
(velocity-addition formula)
$v_{CA}=\frac{v_{BA}+v_{CB}}{1+v_{BA}v_{CB}}$.
\end{corollary}
\begin{proof}
This velocity-addition formula is derived from the equation $L_{CA}=L_{BA}\cdot L_{CB}$.
\end{proof}

\begin{corollary}\label{corollary1}
(``length contraction/extension'' formula)
Given inertial bodies $A$, $B$ and $C$ such that $v_{AC}=v_{BC}$. Let $\Delta x=\left|x_{AA}(\tau_A)-x_{BA}(\tau_A)\right|$ be the distance between $A$ and $B$ in the reference frame $O_A$. Let $\Delta x'=\left|x_{AC}(\tau_C)-x_{BC}(\tau_C)\right|$ be the distance between $A$ and $B$ in the reference frame $O_C$, then $\Delta x'=w_{CA}\cdot \Delta x$.
\end{corollary}
\begin{figure}
\begin{center}
\includegraphics[scale=0.6]{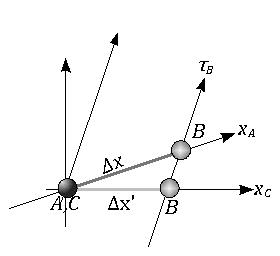}
\caption{Distances between two bodies $A$ and $B$ in Corollary~\ref{corollary1}}
\end{center}
\end{figure}
\begin{proof}
Notice that the values of $\Delta x$ and $\Delta x'$ are constants. Without loss of generality we assume $\tau_{AC}(0)=\tau_{BC}(0)=0$, $x_{AC}(0)=0$, $v_{AC} \geq 0$, $x_{BA} \geq 0$. Then $x_{BA}(\tau_A)\equiv \Delta x$ and $x_{BC}(0)=\Delta x'$. Let $\tau_A$ be such a moment of time that the events $(x_{BC}(0),0)=(\Delta x',0)$ and $(x_{BA}(\tau_A),\tau_A)=(\Delta x,\tau_A)$ are the same. Then the formula $\Delta x'=w_{CA}\cdot \Delta x$ of 
``length contraction'' follows from $L_{CA}\cdot\left(\begin{array}{c} \Delta x' \\ 0\end{array} \right)=\left(\begin{array}{c} \Delta x \\ \tau_A\end{array} \right)$ and Theorem~\ref{theorem1}.
\end{proof}

As it will be seen, from the example at the end of this section, $w_{CA}$  may take on a value which is 
less than 1 as well as more than 1. So it means that in our discrete model 
we have contracting length as well as extending length in respect to 
different inertial frame system.

Now we give a definition of {\bf internal state} of a body. Let for bodies $A$ and $B$ there be a bijection $\phi:A\rightarrow B$ such that for all $b\in A$ elementary bodies $b$ and $\phi(b)$ are isomorphic. 
We say that $A$ at the moment of proper time $\tau_A$ and $B$ at the moment of proper time $\tau_B$ are {\bf affine isomorphic} iff $\{(\phi(b),x_{bA}(\tau_A)|b\in A\}$=$\{(b,x_{bB}(\tau_B)|b\in B\}$. 
\begin{definition}
Two inertial bodies are in the same internal state at some moments of their proper time iff they are affine isomorphic at their respective proper time.
\end{definition}

Internal state of an inertial body does not depend on its spatial velocity in the absolute reference frame. Thus, the external state of an inertial body can be seen as a combination of two components: the spatial velocity of the body and its internal state.

If we now consider the body as an automata-like computational structure, whose states are defined as the internal states, the seemingly natural question whether a body can determine his own absolute velocity is by definition an algorithmically unsolvable problem or a meaningless question. If body states are by definiton the external states, then the same question is substanceless, since the external state always contains information about the absolute velocity.

In order to illustrate the concept of affine isomorphism let us consider bodies $A_1$ and $A_2$ from the Examples~\ref{example1} and~\ref{example2}. This bodies are affine isomorphic. The corresponding transformation of the reference frame $O_2$ of $A_2$ to $O_1$ of $A_1$ is:

\[ \left( \begin{array}{c}
x'  \\
t' \end{array} \right)
=
\left( \begin{array}{cc}
\frac{3}{2} & \frac{1}{2}  \\
\frac{1}{2} & \frac{3}{2} \end{array} \right)
\left( \begin{array}{c}
x \\
t \end{array} \right)
-
\left( \begin{array}{c}
\frac{1}{2}  \\
\frac{1}{2} \end{array} \right).\]

\begin{figure}
\begin{center}
\includegraphics[scale=0.3]{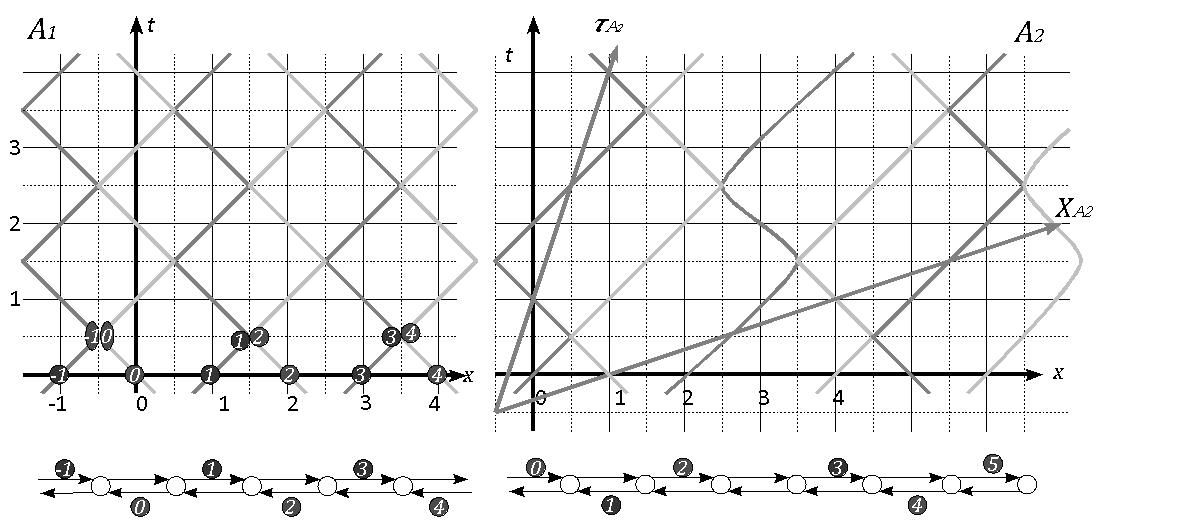}
\caption{The time-space diagrams for the collectives of automata from
Examples~\ref{example1} and~\ref{example2}. }\label{F_ex2}
\end{center}
\end{figure}

The dynamics of the bodies and illustration of the transformation are shown on the Figure~\ref{F_ex2}. From the value of transformation matrix and Corollary~\ref{cor3} it follows that $v_{21}=-v_{12}=1/3$ $w_{21}=2/3$, $w_{12}=4/3$.

\section{Final Remarks}\label{remarks}

Let us compare the obtained results with formulas of special relativity theory.
It is interesting to have a look, from our model viewpoint, at two equations $\Delta t'={\Delta t}/\sqrt{1-(v/c)^2}$ of time dilation and $\Delta x'={\Delta x}\cdot\sqrt{1-(v/c)^2}$ of length contraction of the special relativity theory. Drawing a proper analogy between them and $\tau_{AC}(\tau_C)-\tau_{AC}(0)=w_{AC}\cdot\tau_C$ (Remark~\ref{rem1}) and $\Delta x'=w_{CA}\cdot \Delta x$ (Corollary~\ref{corollary1}) respectively we can see, due to generally asymmetry $w_{AC}\neq w_{CA}$ in our discrete virtual ``world'', that the coefficient $1/\gamma=\sqrt{1-(v/c)^2}$ reciprocal to Lorentz factor $\gamma$ has different ``physical'' meanings in these formulas. The factor $1/\gamma$ has in the first equations a meaning of the coefficient $w_{AC}$ and in the second equations has a meaning of the coefficient $w_{CA}$ if we consider a ``moving'' $A$ with respect to a ``rest'' $C$.

\section{Final Conclusion}\label{conclusion}

Not difficult to generalize the approach developed in this work to the case of finite one-dimensional environments, as well as non-inertial-body case. 
However, we can show that for the case of two-dimensional discrete environment (Cartesian product of two one-dimensional environments) the transformation connecting two inertial reference frames can not be affine in general. 
This follows from the fact that the reference frames are three-dimensional in this case, and affine transformation matrix must have exactly four eigenvectors (there are so many different directions of elementary motions in the two-dimensional discrete enironment by Lemma~\ref{lemma_LBA}). This fact is one of the most interesting dim consequences of this work.

We would like to position this paper from the finite automata theory point of view as an introductory research work on vague 
fundamental notion of a state for distributed computational dynamic structure.\\

{\bf Acknowledgements}: 
The author acknowledges the useful discussions on this work with Dr. Valeriy Kozlovskyy, Dr. Igor Grunsky and Dr. Igor Potapov.





\bibliographystyle{elsarticle-num}



\end{document}